\documentclass{eas}
\usepackage{graphicx}
\usepackage{natbib}
\bibpunct{(}{)}{;}{a}{}{,} 

\begin{document}

\title{IPOPv2: Photoionization of Ni XIV - a test case. }

\author{F. Delahaye} 
\address{LERMA, Observatoire de Paris, ENS, UPMC, UCP, CNRS, 5 Place Jules Janssen, F-92195 Meudon Cedex, France}
\author{P. Palmeri} 
\address{Astrophysique et Spectroscopie, Universit\'e de Mons - UMONS, B-7000 Mons, Belgium}
\author{P. Quinet} 
\sameaddress{2} 
\secondaddress{IPNAS, Universit\'e de Li\`ege, B-4000 Li\`ege, Belgium}
\author{C. J. Zeippen} 
\sameaddress{1}

\begin{abstract}
Several years ago, M. Asplund and coauthors (2004) proposed a revision
of the Solar composition. The use of this new prescription for Solar
abundances in standard stellar models generated a strong disagreement
between the predictions and the observations of Solar observables.
Many claimed that the Standard Solar Model (SSM) was faulty, and more
specifically the opacities used in such models. As a result, activities
around the stellar opacities were boosted. New experiments (J. Bailey
at Sandia on Z-Pinch, The OPAC consortium at LULI200) to measure
directly absorbtion coefficients have been realized or are underway.
Several theoretical groups (CEA-OPAS, Los Alamos Nat. Lab., CEA-SCORCG,
The Opacity Project - The Iron Project (IPOPv2) ) have started new sets
of calculations using different approaches and codes. While the new
results seem to confirm the good quality of the opacities used in SSM,
it remains important to improve and complement the data currently
available.  We present recent results in the case of the photoionization
cross sections for Ni XIV ($Ni^{13+}$) from IPOPv2 and possible implications on
stellar modelling.
\end{abstract}
\maketitle
\runningtitle{Photoionization of NiXIV - RMATRIX vs AS}

\section{Introduction}

Recent studies on the solar abundances of elements have questioned the accuracy of the available sets of
opacities. \citet{Asplundetal2004} proposed an abundance for oxygen which was much reduced as
compared to previous values. Decreased values for the C and N abundances were also recommended.
One of the consequences was to upset seismology findings so far considered reliable (depth of the convection zone,
He content at the surface, sound speed profile) and one suggested way out of this difficulty was to
modify the stellar opacities. Even though the revisited abundances have increased again since 2004,
the debate is going on about the reliability of the opacities used by the astrophysical community and
about the relevance of the ``new solar abundances'' \citep[see for example][DP06 hereafter]{PinsonneaultDelahaye2009, 
Delahayeetal2010, Nahar2011, DelahayePinsonneault2006}.  

This debate ignited several efforts to quantify the uncertainties on the opacities. 
Experimental work is under way using powerful lasers at LULI2000 \citep{Gilles2011}
or Z-pinch machine \citep{Bailey2009} in order to measure opacities.
Dedicated theoretical work and systematic comparisons between different findings
for conditions characterizing the solar structure have been started within the OPAC consortium.
Some results and comparisons for Fe and Ni are already available
\citep[see for example][]{TurckChiezeetal2012}.

As presented in \cite{Badnelletal2005} or DP06, the mean opacities from the OP for mixtures
like the solar case agree well with other work like OPAL \citep{OPAL1992} and LEDCOP (Magee et al. 1995). 
More recent calculations from the OPAS group \citep[B12 hereafter]{Blancard2012} also agree very well. 
Also, it is worth mentioning that modeled helioseismic tests have been performed with LEDCOP opacities
(\citet{NeuforgeVerheeckeetal2001}) as well as with OPAL and OP data (DP06), showing good agreement
in the prediction of solar observables. This gives us a pretty good confidence in the quality of the
Rosseland mean opacities provided for solar models. However, as shown in \citet{Bailey2007},
the details of the monochromatic opacities for certain elements may show sizeable differences
with experimental data.  \citet{Delahaye2005} and The OPAS group (B12) also provided detailed
comparisons  for individual elements showing large differences. While these differences will not affect
our results on the solar abundances presented in DP06, it is essential to
quantify and understand them. The effect of remaining discrepancies could be of more
consequence when running an evolutionary code. Discrepancies on spectral opacities for
individual elements will modify the velocity of microdiffusion processes which in turn
will influence the structure and evolution of intermediate mass stars \citep{Richard2001,Delahaye2005}.
It also will impact the condition of pulsating stars as shown in \citet{MontalbanMiglio2008}.

The present work present partial results for the Ni atomic data. The goal is to evaluate the two different
approaches ($R$-matrix and independent-processes, isolated-resonance, distorted-wave (IPIRDW)) to calculate photoionization cross sections for the Ni~{\sc xiv} ion and estimate the impact on the opacities. In the next section we present the strategy and highlight the main differences on two approaches used to calculate the cross sections . In section 3 we describe the calculations of cross sections in both methods. 
In section 4 the results are presented comparing the different approaches and estimate the consequences on the opacities. We present the conclusion and perspective in the last section.

\section{Strategy}

Opacities are a property of a medium characterizing its resistance to the transfer of light.
In the dertermination of such absorbtion coefficients,
one has to take into account different processes, photo-excitation, scattering and photoionization .
For each element at each frequency point, the combination of these processes
gives the total contribution of the element (see Eq. 2.1)
\begin{equation}
\kappa_{tot}(\nu) = \kappa_{bound-bound}(\nu) + \kappa_{bound-free}(\nu) + \kappa_{free-free}(\nu)
\end{equation}

In order to derive the Rosseland mean, we have to sum up these contributions.
Depending on the element and the physical conditions, the contribution of each
processe will vary. In the case of stellar interiors and specificaly in the case of the sun,
for all metals, the contribution from the photionization is dominant as shown in B12.
In our project to quantify all uncertainties in the process of calculating opacities, it is then a
good starting point to look at the photoionization cross sections. In the present
work we look at the effect of the different methods to calculate the
photoionization absorption cross section and the impact on the derived opacities.

For the conditions we are interested in, they are two main ways to photoionize an atom.
There is the direct process where one photon is absorbed and takes one electron
away from the atom to make it free (see Eq. 2.2 ) and there is the indirect path where
the photon excites one electron into an autoionizing state (also called quasi bound state)
which autoionizes faster than it decays radiatively (see Eq. 2.3). This is illustrated
in Figure~1.

\begin{figure}[h]
   \includegraphics[scale=0.4]{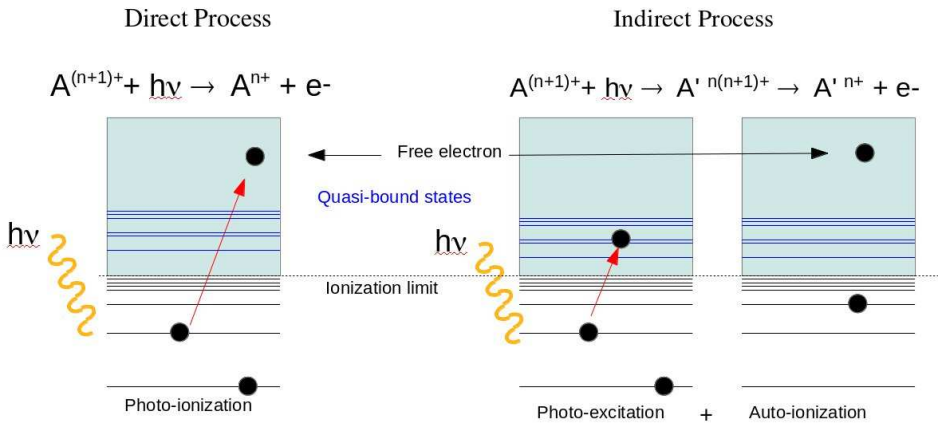}
   \caption{\label{Figure 1} Photoionization: 1 or 2 steps processes}
\end{figure}

Direct process:
\begin{equation}
A^{N+1} _{i} + h\nu  \rightarrow A^{N} _{f} + e^-
\end{equation}

Indirect process (also called photoexcitation-autoionization):
\begin{equation}
A^{N+1}_{i} + h\nu  \rightarrow A^{N+1} _{j} \rightarrow A^{N} _{f} + e^-
\end{equation}
Both processes can interfere and give rise to resonances that 
display Beutler-Fano profiles \citep{Fano1961} in the cross sections. 

The quality of OP opacities, especialy when looking at monochomatic opacity 
details for certain elements (Fe, Ni) \citep{Bailey2009, Nahar2011, TurckChiezeetal2012},
can legitimately be questionned, even if it has been shown that the Rosseland 
mean opacities of a solar mixture are in very good agreement (DP06,B12).
Moreover, the spectral opacities provided by OP include Mn, Cr and Ni. 
However the atomic data for these elements are not available. 
Their spectral opacities have been determined using Fe
atomic data via a scaling procedure.

The goal of the present work is to estimate the impact of the different
approximations and different methods on the atomic data and on the spectral
opacities in detail. We want to quantify the uncertainties due to such
choices ($R$-matrix vs IPIRDW, direct calculations vs scaling, LS
coupling vs intermediate coupling) on opacities and estimate the impact
on application after propagation of these uncertainties.

More specifically, in the present work, depending on the methods used to calculate
the corresponding cross sections, the treatment of resonances will be drastically different.
In the $R$-matrix method, both path for photoionization are included
naturaly and are interfering, giving rise to Beutler-Fano resonances.
In the IPIRDW method on the other hand, the inteference phenomena is neglected. 
Only lines appear on top of the continuum.
A priori it would be more appealing to choose the more complete methods and be more
accurate. However in calcuating opacities it is always a question of balance between
completeness and accuracy. $R$-matrix calculations are far more demanding in order to
include a large enough target and do not allow any treament of the broadening of
the lines (resonances) while IPIRDW calculations allow to extend at will the number of lines
included to take into account the resonances and it is possible to apply a profile
for the broadening.  

\section{Photoionization cross section calculations}

\subsection{$R$-matrix method}

One set of photoionization cross sections in this work has been calculated using the Breit-Pauli 
$R$-matrix method \citep[e.g.][]{Berrington1995} as employed in the Iron Project (IP) and utilized in a number of previous
publications. The aims and methods of the IP are presented in \citet{Hummer1993}. We
briefly summarize the main features of the method and calculations.
In the close coupling (CC) approximation the wavefunction expansion, $\Psi(E)$ , for a
total spin and angular symmetry $S L \pi$ or $J \pi$, of the (N + 1) electron system is
represented in terms of the target ion states as
$$\Psi (E) = A \sum_{i} \chi _i \Theta _i + \sum_{j} c_j \Phi _j$$
where $\chi _i$ is the target ion wavefunction in a specific state $S_i L_i \pi _i$ or level $J_i \pi _i$ , and $\Theta _i$ is the
wavefunction for the (N + 1)th electron in a channel labelled as $S_i L i_ (J_i )\pi _ii k_i ^2 l_i (S L\pi) [ J \pi]$;
$k_i ^2$  is the incident kinetic energy. In the second sum the $\Phi _j$ are correlation wavefunctions of
the (N + 1) electron system that (a) compensate for the orthogonality conditions between the
continuum and the bound orbitals, and (b) represent additional short-range correlations that are
often of crucial importance in scattering and radiative CC calculations for each symmetry. The
$\Phi _j$ are also referred to as bound channels, as opposed to the continuum or free channels
in the first sum over the target states. In the Breit-Pauli $R$-matrix calculations the set of $S L \pi$ are
recoupled in an intermediate (pair) coupling scheme to obtain (e + ion) states with total $J \pi$,
followed by diagonalization of the (N +1)-electron Hamiltonian.

In the present work the target expansion for the CC calculations consists of 23 $LS$ terms with principal
quantum number up to $n = 4$ giving rise to 812 levels. The target eigenfunctions were developed using the {\sc autostructure}
program \citep{Badnell2011} an extension of {\sc superstructure} \citep{Eissner1974,NussbaumerStorey1978}.

The full expansion is 3s$^2$3p$^2$ + 3s3p$^3$ + 3s$^2$3p3d + 3s3p$^2$3d +
3p$^4$ + 3p$^3$3d + 3s$^2$3d + 3p$^2$3d$^2$ + 3s3p3d$^2$ + 3p3d$^3$ + 3s3d$^3$
+ 3s$^2$3p4$l$ ($l$=0$-$3) + 3s3p$^2$4$l$ ($l$=0$-$3) + 3p$^3$4$l$ ($l$=0$-$3).
with the scaling factors, $ \lambda_{1s} =  1.40434\ ,   
\lambda_{2s} = 1.12157\ ,
\lambda_{2p} = 1.06304\ ,
\lambda_{3s} = 1.13180\ ,
\lambda_{3p} = 1.09963\ ,
\lambda_{3d} = 1.12434\ ,
\lambda_{4s} = 1.15189\ ,
\lambda_{4p} = 1.11693\ ,
\lambda_{4d} = 1.14682\ ,
\lambda_{4f} = 1.29663 $ in the Thomas-Fermi-Dirac-Amaldi (TFDA)
potential, $V(\lambda_{nl})$, employed in {\sc autostructure}. 
This target is used in both methods ($R$-matrix and IPIRDW) in order to minimise
the differences between the two runs in order to determine precisely the effect of each ingredient,
in this case the configuration interaction on the resonance part of the cross sections.
                         
In order to estimate the quality of the target wavefunction expansion, we compare the
energy levels and the transition probabilities with those from the National Institute 
for Standards and Technology \citep{NIST_ASD,NISTlevels,Fuhr1988}. The errors are of the order of 3.8\%
maximum for the energy levels and within 15\% for all lines but one for which the error reach 30\%. 
However it is difficult to rate the level of agreement since the stated NIST accuracy is no better than
50\% for these transitions.

\subsection{IPIRDW method}

The photoionization cross sections of all the bound states of Ni$^{13+}$ up to $nl=10g$ have
been computed using the atomic structure code {\sc autostructure}.
It computes term energies and/or fine-structure level energies, radiative and Auger rates,
photoionization cross sections and electron-impact collision strengths.
In these calculations, the atomic orbitals are constructed by diagonalizing
the non-relativistic Hamiltonian within a scaled statistical TFDA
model potential. The scaling parameters
are optimized variationnally by minimizing a weighted sum of the LS term energies.
$LS$ terms are represented by configuration-interaction (CI) wavefunctions of the type
\begin{equation}
\Psi(LS)= \sum_i~c_i~\Phi_i
\end{equation}
Continuum wavefunctions are constructed within the distorted-wave (DW) approximation.
Relativistic fine-structure levels and rates can be obtained by diagonalizing the Breit-Pauli
Hamiltonian in intermediate coupling. The one- and two-body operators $-$ fine-structure
and non-fine structure $-$ have been fully implemented to order $\alpha^2 Z^4$ where $\alpha$
is the fine-structure constant and $Z$ the atomic number.

The non-resonant direct photoionization process (DPI, Eq.~2.2) has been treated separately 
from the photoexcitation-autoionization (PAI, Eq.~2.3) process within the framework of the 
IPIRDW approximation \citep{BadnellSeaton2003,Fu2008}. 

In order to compare with our $R$-matrix calculation, we have computed the
cross sections in LS coupling  and used the same interacting configurations to describe the target of
the scattering problem. The scaling parameters for the target orbitals, i.e. 1s to 4f, 
were identical to our $R$-matrix model.
Relativistic corrections have been included in the orbital calculations by adding to the
radial equations the mass-velocity and the Darwin terms.

Separate calculations have been performed for each initial complex of  Ni$^{13+}$. For the $n=3,4$
complex we have considered the following DPI and PAI channels
\begin{equation}
\{3l^4,3l^34l\}\{3l,4l\} + h\nu \rightarrow \{3l^4,3l^34l\} + e^- (kl'')
\end{equation}
\begin{equation}
\begin{array}{lcl}
\{3l^4,3l^34l\}\{3l,4l\} + h\nu  & \rightarrow  & \{3l^4,3l^34l\} n'l'  (n' > 4 , l' \le 5)   \\
  & \rightarrow  &  \{3l^4,3l^34l\} + e^- (kl'') 
\end{array}
\end{equation}

\noindent where the initial $n=3,4$ complex is built by adding a $3l$ or a $4l$ electron to the target 
complex  $\{3l^4,3l^34l\}$. Similarly the DPI and PAI channels considered for the other initial 
complexes (up to $nl'=10g$)  were
\begin{equation}
\{3l^4,3l^34l\}nl'  + h\nu \rightarrow \{3l^4,3l^34l\} + e^- (kl''') 
\end{equation}

\begin{equation}
\begin{array}{lcl}
\{3l^4,3l^34l\}nl'  + h\nu & \rightarrow &\{3l^4,3l^34l\} n'l''  (n' > n, l'' \le 5)\\
& \rightarrow & \{3l^4,3l^34l\} + e^- (kl''') 
\end{array}
\end{equation}

The scaling parameters of the Rydberg (other than the target orbitals) and
continuum orbitals have been fixed to 1.23. This value has been obtained
optimizing the scaling parameter of the Rydberg orbital $69s$ by minimizing
all the 588 terms of the $\{3l^4,3l^34l\}69s$ complex where the other orbitals
were fixed to their target values.

A second IPIRDW calculation has been carried out to
obtain the photoionization cross section of the Ni$^{13+}$ ground state
using the orbitals optimized in the initial atomic system of the photoionization
process. This was done in order to test the influence on the cross sections of
the choice of atomic orbitals. The scaling parameters of the closed shells
(1s to 2p) have been fixed to 1 and those of the $n=3,4$ orbitals have been
optimized minimizing all the 4964 terms of the Ni$^{13+}$ complex  $\{3l^4,3l^34l\}\{3l,4l\}$
( $ \lambda_{1s\ to\ 2p} =  1.00\ ,   
\lambda_{3s} = 1.26\ ,
\lambda_{3p} = 1.18\ ,
\lambda_{3d} = 1.18\ ,
\lambda_{4s} = 1.23\ ,
\lambda_{4p} = 1.16\ ,
\lambda_{4d} = 1.15\ ,
\lambda_{4f} = 1.27,\ $ ).
The scaling parameters of the other orbitals (Rydberg and continuum)
have been fixed to 1.25. The latter has been obtained in a similar procedure
as used in the first IPIRDW calculation but fixing the $n=3,4$ orbitals to those of the initial atomic
system. The DPI and PAI channels considered were identical to those shown in Eqs.~$(3.2-3.3)$.

\section{Results and discussion} 

The cross section for the ground state of Ni~{\sc xiv} from our $R$-matrix calculation
are plotted (in black) along with the results of the first IPIRDW calculation (in red)
in Figure 2. We can remark the importance of the interference phenomena, while the lines 
added to the continuum in the IPIRDW results are all above the continuum
with a symetric shape (we used a Lorentzian profile for the resonances), the structure 
in the $R$-matrix results are more complex. The interaction
between the DPI and PAI channels translate into a specific asymetric Beutler-Fano profile. Hence the black plot
shows features going below the continuum which correspond to a reduction of the absorption coefficient
at these specific frequencies and a 'large' wing above the continuum redward to the center of the line.

Differences between our first (black) and second IPIRDW (red) cross sections can be seen in Figure~3.
The DPI (continuum) part of our second IPIRDW calculation that uses the orbitals optimized 
in the initial $(N+1)$-system is systematically overestimated with respect to our first IPIRDW model. 
Even the positions of the resonances and the thresholds differ significantly. Our first IPIRDW model is 
clearly the best choice in regard to the good agreement with our $R$-matrix model (see Figure~$2$).

Several steps are required to calculate opacities. The first one is the determination
of ionic fractions and level populations for each atom present in the plasmas.
It depends on the  physical conditions (T, $\rho$). Then given the populations
one can sum the bound-bound, bound-free, free-free and scattering to obtain the opacities.
We do not have all the data here to provide the full results but as an exercice
we consider a plasma with only one ion, Ni~{xiv}, populated only on the ground state.
We can then quantify the effect of the interference in the opacities.

For the spectral opacities we simply have the same signature as in the photoionization cross sections. 
It is then interesting to evaluate some mean opacities to see the effect of the Fano profiles. 
It is only a toy model since the temperature variation along with the
density change would modify the ionic fraction as well as the populations. In the present case it is only supposed to
give a first estimate on the specifics of the atomic codes. In the present case, we first omitted the
weighting function in the integrand of the Rosseland and Planck mean opacity calculation. We used the following
formulae for the Rosseland ($\kappa_R$) and Planck ($\kappa_P$) mean approximation respectively.

$$
\begin{array}{lr}
	\frac{1}{\tilde \kappa_R ^{bf}} = \int _0 ^{\nu_{max}} \frac{1}{\kappa_{bf}} (\nu) d \nu \ \ and\ &
    \tilde \kappa_P ^{bf} = \int _0 ^{\nu_{max}} \kappa_{bf} (\nu) ~d \nu 
\end{array}
$$

In Table~1, the relative differences between 'Rosseland' and 'Planck' mean opacities with 
respect to values computed using our $R$-matrix model are reported for our
two different IPIRDW cross sections and for our $R$-matrix cross section where the  
photon energy resolution has been degraded ($R$-matrix (low resolution)).
One can see that the Rosseland mean is more sensitive to details of the cross sections than
the Planck mean and that mean opacities based on our first IPIRDW model better agree 
with those based on our accurate $R$-matrix model even with respect to the ones based on
a low-resolution $R$-matrix model. The differences are of the order of the usually expected
accuracies of the photoionization cross sections ($\sim$20\%).

\section{Conclusion and perspectives}

Our preliminary results on the photoionization of Ni$^{13+}$ ground state shows the 
impact of the different methods on the cross section as well as some estimated
mean opacities. These differences are important given the required accuracy 
for application. However it is too early to draw definitive conclusion since
other aspect of the calculation of opacities have not yet been included 
(equation of state for the population and ionic fractions, plamas conditions, etc.)
Some cancellation effects can occur. The full calculation is underway, including 
fine structure and other ions to provide a full picture of the Ni opacities.

\begin{figure}[h]
   \centering
   \caption{\label{Figure 2} Details of the total photoionization cross section for the ground state: 
   comparison between $R$-matrix (in black) and our first IPIRDW calculation (in red).}
   \includegraphics[height=5.0cm,width=11.0cm]{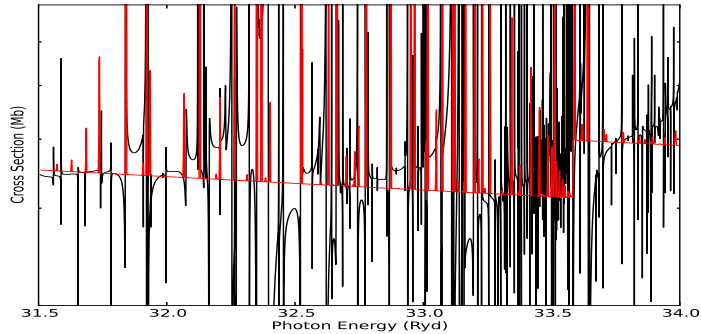}
\end{figure}

\begin{table}[h]
\caption{\% difference, $\Delta\tilde\kappa = (\tilde\kappa - \tilde\kappa(R-{\rm matrix}))/\tilde\kappa(R-{\rm matrix})$, 
in the Rosseland and Planck estimate with
respect to ref (high res. $R$-Matrix). }
\begin{center}
\begin{tabular}{ccc}
\hline
Methods  &  $\Delta\tilde\kappa_P$  &  $\Delta\tilde\kappa_R$\\
\hline
IPIRDW(first)        &       -0.6\%   &  -9.6\%  \\
IPIRDW(second)     &      11.3\%    &  -23.4\%  \\
$R$-matrix (low resolution)   & -6.5\%   &  -14.2\% \\
\hline
\end{tabular}
\end{center}
\end{table}%

\begin{figure}[h]
   \centering
   \includegraphics[scale=0.35,clip=true]{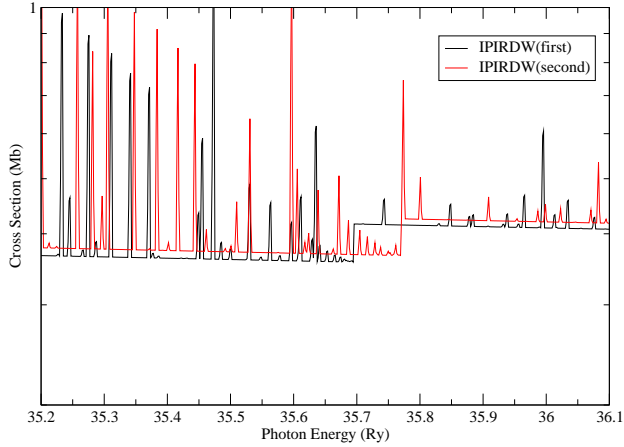}
   \caption{\label{Figure 3} Total photoionization cross section for the ground state:
    comparison between our first (in black) and our second IPIRDW calculation (in red). }
\end{figure}

\bibliography{mnemonic,delahaye_f_Roscoff2013}

\begin{thebibliography}{}

\bibitem[\protect\astroncite{Asplund et~al.}{2004}]{Asplundetal2004}
Asplund, M., Grevesse, N., Sauval, A.~J., Allende~Prieto, C., and Kiselman, D.:
  2004,
\newblock {\bf 417}, 751

\bibitem[\protect\astroncite{Badnell}{2011}]{Badnell2011}
Badnell, N.~R.: 2011,
\newblock {\em Computer Physics Communications} {\bf 182}, 1528

\bibitem[\protect\astroncite{Badnell et~al.}{2005}]{Badnelletal2005}
Badnell, N.~R., Bautista, M.~A., Butler, K., Delahaye, F., Mendoza, C.,
  Palmeri, P., Zeippen, C.~J., and Seaton, M.~J.: 2005,
\newblock {\bf 360}, 458

\bibitem[\protect\astroncite{Badnell and Seaton}{2003}]{BadnellSeaton2003}
Badnell, N.~R. and Seaton, M.~J.: 2003,
\newblock {\em Journal of Physics B Atomic Molecular Physics} {\bf 36}, 4367

\bibitem[\protect\astroncite{Bailey et~al.}{2007}]{Bailey2007}
Bailey, J.~E., Rochau, G.~A., Iglesias, C.~A., Abdallah, Jr., J., Macfarlane,
  J.~J., Golovkin, I., Wang, P., Mancini, R.~C., Lake, P.~W., Moore, T.~C.,
  Bump, M., Garcia, O., and Mazevet, S.: 2007,
\newblock {\em Physical Review Letters} {\bf 99(26)}, 265002

\bibitem[\protect\astroncite{Bailey et~al.}{2009}]{Bailey2009}
Bailey, J.~E., Rochau, G.~A., Mancini, R.~C., Iglesias, C.~A., Macfarlane,
  J.~J., Golovkin, I.~E., Blancard, C., Cosse, P., and Faussurier, G.: 2009,
\newblock {\em Physics of Plasmas} {\bf 16(5)}, 058101

\bibitem[\protect\astroncite{Berrington et~al.}{1995}]{Berrington1995}
Berrington, K.~A., Eissner, W.~B., and Norrington, P.~H.: 1995,
\newblock {\em Computer Physics Communications} {\bf 92}, 290

\bibitem[\protect\astroncite{Blancard et~al.}{2012}]{Blancard2012}
Blancard, C., Coss\'e, P., and Faussurier, G.: 2012,
\newblock {\bf 745}, 10

\bibitem[\protect\astroncite{Delahaye and Pinsonneault}{2005}]{Delahaye2005}
Delahaye, F. and Pinsonneault, M.: 2005,
\newblock {\bf 625}, 563

\bibitem[\protect\astroncite{Delahaye and
  Pinsonneault}{2006}]{DelahayePinsonneault2006}
Delahaye, F. and Pinsonneault, M.~H.: 2006,
\newblock {\bf 649}, 529

\bibitem[\protect\astroncite{Delahaye et~al.}{2010}]{Delahayeetal2010}
Delahaye, F., Pinsonneault, M.~H., Pinsonneault, L., and Zeippen, C.~J.: 2010,
\newblock {\em ArXiv e-prints}

\bibitem[\protect\astroncite{Eissner et~al.}{1974}]{Eissner1974}
Eissner, W., Jones, M., and Nussbaumer, H.: 1974,
\newblock {\em Computer Physics Communications} {\bf 8}, 270

\bibitem[\protect\astroncite{Fano}{1961}]{Fano1961}
Fano, H.: 1961,
\newblock {\em Phys. Rev. A} {\bf 124}, 1866

\bibitem[\protect\astroncite{Fu et~al.}{}]{Fu2008}
Fu, J., Gorczyca, T.~W., Nikolic, D., Badnell, N.~R., Savin, D.~W., and Gu,
  M.~F.,
\newblock {\bf 77(3)}, 032713

\bibitem[\protect\astroncite{Fuhr et~al.}{1988}]{Fuhr1988}
Fuhr, J.~R., Martin, G.~A., and Wiese, W.~L.: 1988,
\newblock {\em Journal of Physical and Chemical Reference Data} 17

\bibitem[\protect\astroncite{Gilles et~al.}{2011}]{Gilles2011}
Gilles, D., Turck-Chi\`eze, S., Loisel, G., Piau, L., Ducret, J.-E., Poirier,
  M., Blenski, T., Thais, F., Blancard, C., Coss\'e, P., Faussurier, G.,
  Gilleron, F., Pain, J.~C., Porcherot, Q., Guzik, J.~A., Kilcrease, D.~P.,
  Magee, N.~H., Harris, J., Busquet, M., Delahaye, F., Zeippen, C.~J., and
  Bastiani-Ceccotti, S.: 2011,
\newblock {\em High Energy Density Physics} {\bf 7}, 312

\bibitem[\protect\astroncite{Hummer et~al.}{1993}]{Hummer1993}
Hummer, D.~G., Berrington, K.~A., Eissner, W., Pradhan, A.~K., Saraph, H.~E.,
  and Tully, J.~A.: 1993,
\newblock {\bf 279}, 298

\bibitem[\protect\astroncite{Kramida et~al.}{2013}]{NIST_ASD}
Kramida, A., Ralchenko, Y., Reader, J., , and Team, N.~A.: 2013,
\newblock NIST Atomic Spectra Database (ver. 5.1), [Online]. Available
  thttp://physics.nist.gov/asd [2013, September 19]. National Institute of
  Standards and Technology, Gaithersburg, MD.

\bibitem[\protect\astroncite{Montalban and Miglio}{2008}]{MontalbanMiglio2008}
Montalban, J. and Miglio, A.: 2008,
\newblock {\em Communications in Asteroseismology} {\bf 157}, 160

\bibitem[\protect\astroncite{Nahar et~al.}{2011}]{Nahar2011}
Nahar, S.~N., Pradhan, A.~K., Chen, G.-X., and Eissner, W.: 2011,
\newblock {\bf 83(5)}, 053417

\bibitem[\protect\astroncite{Neuforge-Verheecke
  et~al.}{2001}]{NeuforgeVerheeckeetal2001}
Neuforge-Verheecke, C., Guzik, J.~A., Keady, J.~J., Magee, N.~H., Bradley,
  P.~A., and Noels, A.: 2001,
\newblock {\bf 561}, 450

\bibitem[\protect\astroncite{Nussbaumer and
  Storey}{1978}]{NussbaumerStorey1978}
Nussbaumer, H. and Storey, P.~J.: 1978,
\newblock {\bf 64}, 139

\bibitem[\protect\astroncite{Pinsonneault and
  Delahaye}{2009}]{PinsonneaultDelahaye2009}
Pinsonneault, M.~H. and Delahaye, F.: 2009,
\newblock {\bf 704}, 1174

\bibitem[\protect\astroncite{Richard et~al.}{2001}]{Richard2001}
Richard, O., Michaud, G., and Richer, J.: 2001,
\newblock 558

\bibitem[\protect\astroncite{Rogers and Iglesias}{1992}]{OPAL1992}
Rogers, F.~J. and Iglesias, C.~A.: 1992,
\newblock {\bf 401}, 361

\bibitem[\protect\astroncite{Shirai et~al.}{2000}]{NISTlevels}
Shirai, T., Sugar, J., Musgrove, A., and Wiese, W.~L.: 2000,
\newblock {\em Spectral Data for Highly Ionized Atoms: Ti, V, Cr, Mn, Fe, Co,
  Ni, Cu, Kr, and Mo}

\bibitem[\protect\astroncite{Turck-Chi\`eze et~al.}{2012}]{TurckChiezeetal2012}
Turck-Chi\`eze, S., Loisel, G., Gilles, D., Ducret, J.-E., and et~al.: 2012,
\newblock in H. Shibahashi, M. Takata, and A.~E. Lynas-Gray (eds.), {\em
  Progress in Solar-Stellar Physics with Helio- and Asteroseismology}, Vol. 462
  of {\em Astronomical Society of the Pacific Conference Series}, p.~95

\end{thebibliography}
\bibliographystyle{astron}

\end{document}